# CONSTRUCTING AN EFFICIENT MACHINE LEARNING MODEL FOR TORNADO PREDICTION

Editors of the Series WP7
"Mathematical methods for decision making
in economics, business and politics"
*Aleskerov Fuad, Mirkin Boris, Podinovskiy Vladislav*

Tornado prediction methods and main mechanisms of tornado genesis were analyzed. A model, based on the superposition principle, has been built. For efficiency evaluation, the constructed model has been tested on real-life data obtained from the University of Oklahoma (USA).

It is shown that the constructed tornado prediction model is more efficient than all previous models.

*Fuad Aleskerov* – National Research University Higher School of Economics, V.A. Trapeznikov Institute of Control Sciences of Russian Academy of Sciences, Moscow, Russia.

*Nikita Baiborodov* – Moscow Institute of Physics and Technology, Moscow, Russia.

*Sergey Demin* – National Research University Higher School of Economics, Moscow, Russia.

*Sergey Shvydun* – National Research University Higher School of Economics, V.A. Trapeznikov Institute of Control Sciences of Russian Academy of Sciences, Moscow, Russia.

*Theodore Trafalis* – University of Oklahoma, USA.

*Michael Richman* – University of Oklahoma, USA.

*Vyacheslav Yakuba* – National Research University Higher School of Economics, V.A. Trapeznikov Institute of Control Sciences of Russian Academy of Sciences, Moscow, Russia.

# 1. Introduction

Tornadoes are a major disaster hazard. For instance, direct losses to the US economy, caused by tornadoes for the period from 2010 to 2014 were about 16.5 billion dollars (NOAA, US National Weather Service). Therefore, accurately forecasting these events as far in advance as possible is an important research activity.

However, achieving forecasts with long lead times is difficult because tornado genesis process is still not understood fully [Rasmussen et al., 2004]. Furthermore, it has been noted that a crucial part of the process is chaotic, which makes prediction of tornado even more complicated [Rhodes, Senkbeil, 2014].

The approach to model tornadogenesis using dynamic processes is time and resource consuming, and owing to the small scale of a tornado, physically-based numerical weather prediction models still do not provide very good results.

In this work we propose a new approach into disaster prediction by using the methods of intelligent data analysis for detecting tornadic circulations from the set of all observations $A$. As a result, constructed models predict tornado occurrence with higher efficiency. In addition, intelligent data analysis methods work faster than simulation by dynamic weather prediction models, allowing extra time for reacting and preventing dangerous repercussions of the disaster.

The basic goal here is to find the main patterns between different characteristics of air circulation and, given these patterns, detect tornadoes. For this purpose, the following parameters of air circulations are used – temperature, air pressure, relative humidity, velocity of the air flow and many other physical characteristics of the near-storm atmosphere.

There are numerous procedures that allow choosing alternatives from the initial set. In this work, we consider choice procedures of a special type based on the superposition principle[1] [Aleskerov et al., 2014].

Superposition has several advantages. First, the computational complexity of the model can be managed. Unfortunately, most existing machine learning algorithms have a high computational complexity, so they cannot be applied in the case of large number of observations or/and criteria. The use of superposition allows reducing the complexity by applying methods with a low computational complexity on first stages and more accurate methods on final stages. Consequently, our models can be applied to larger initial datasets.

---

[1] In some works the term "composition" is used instead of "superposition".

Second, superposition allows us to combine different methods and use several criteria simultaneously on each step. Moreover, models based on the idea of superposition can be interpreted easily since we can apply several simple methods instead of a complicated one. In addition, superposition will help to reduce the influence of drawbacks of initial methods. Hence, our model may have advantages of all previous techniques.

**Acknowledgement**

The article was prepared within the framework of the Basic Research Program at the National Research University Higher School of Economics (HSE) and supported within the framework of a subsidy by the Russian Academic Excellence Project "5–100".

## 2. Literature review

We consider here only the studies on application of intelligent data analysis methods to the tornado prediction.

Adrianto et al. [2009] proposed the use of support vector machines (SVM) for tornado prediction, i.e. the algorithm constructs a hyperplane that separates a set of elements into two classes (e.g., tornadoes and non-tornadoes).

The defining function is presented as a dot product of two vectors. The first one is a vector of circulation attributes, such as wind speed, temperature, air pressure on the surface, relative humidity at different levels, etc. In turn, the second vector consists of weight coefficients, which show the importance of each parameter in tornado prediction process. So, small absolute value of weight coefficient means that the parameter almost cannot influence future tornado detection. Correspondingly, high value of the coefficient indicates high prediction power of the attribute.

However, the accuracy of all methods based on linear regression is very low. Recently, Trafalis et al. [2014] also proposed other approaches to this problem.

The first mentioned technique is an improvement of the aforementioned method of support vector machines with reverse features elimination (SVM-RFE). According to this approach, one should realize the standard SVM algorithm and get weight coefficients. Afterwards the attribute which has the lowest value of the weight coefficient is eliminated from the features list. Subsequently, the algorithm iteratively continues until only one parameter remains.

At the end of the process, the sequence of eliminated attributes transforms into the ranking of the parameters by the following rule. The number of the attribute in the ranking is equal to the number of iterations (backward), when the attribute was eliminated.

However, the ranking is not the only goal of this method. Some features of air circulation might not help, but even interfere with tornado detection (giving false signals). Therefore, the last step of the SVM-RFE algorithm is choosing the number of parameters which are used for tornado detection. For this purpose, one evaluates the accuracy of the method according to the number of used attributes (it is important to point out that the algorithm uses only top attributes from the ranking).

One more method proposed for tornado prediction is a neural network approach [Marzban et al., 1996, Marzban, 2000, Lakshmanan et al., 2005]. The main idea is the construction of a non-linear function that maps real-valued input variables to a number varying from 0 to 1. The most popular version of neural network is used – three-layer perceptron network. Such a neural network has one hidden layer of neurons.

The next proposed technique is Random Forest method (RANF), developed by Breiman [2001]. It is a group classifier with multiple decision trees, wherein each tree is trained on a part of all attributes, chosen randomly. Consequently, the predictions of all trees are aggregated by the classification based on the majority of the votes over all classifiers.

The last proposed algorithm is Rotation Forest method (ROTF) by Rodriguez et al. [2006]. This method utilizes principal component analysis (PCA) to distinguish attributes which are used to construct a decision forest. This forest consists of decision trees which are trained on the whole data set in a rotated feature space. The set of parameters is randomly divided into K groups. Afterwards, principal component analysis is performed simultaneously on each group. As a result, we receive K classifiers that compose one classification instrument. This instrument works in the same way as RANF; classification is based on the majority of the votes over all of the trees.

It is important to point out that all the aforementioned methods have specific drawbacks, which decrease the accuracy of tornado prediction. Accord-ing to the aforementioned previous studies, all techniques detect only a fraction of the tornadoes (more than 25% of all tornadoes are missed). In addition, a significant part of tornado signals (about 20%) is false alarm, which is also very important. Thus, general accuracy of these methods (from 51 to 57%) is not high enough for efficient application in real life, because such results slightly exceed tornado prediction by the coin toss.

In this work we use choice functions, based on the superposition principle [Aleskerov et al., 2014] to make tornado prediction. Usually a standard choice function $C_1(\cdot)$ consists in the choice of some subset of alternatives that satisfy the predefined condition.

Ideally the specified condition should be "being tornadic circulation." However, in real life it is difficult to satisfy this condition. Hence, we use some simpler conditions, which narrow the initial set of observations and get trustworthy results by application of the superposition principle.

Superposition of two choice functions $C_1(\cdot)$ and $C_2(\cdot)$ is a binary operation $\odot$, the result of which is a new function $C^*(\cdot) = C_2(\cdot) \odot C_1(\cdot)$, which has a form $\forall X \in 2^A C^*(X) = C_2(C_1(X))$ [Aizerman, Aleskerov, 1995]. In short, the latter function $C_2(\cdot)$ is used on the data obtained by the application of the former one $C_1(\cdot)$. It is necessary to mention that in the case of change of methods' application order the result might be totally different, as the superposition is not commutative and the functions $C_1(\cdot)$ and $C_2(\cdot)$ can be completely diverse. The properties of the superposition operator were studied in [Aizerman, Aleskerov, 1995; Aleskerov, Cinar, 2008; Shvydun, 2015] and other papers.

## 3. Preliminary data analysis

### 3.1. Data description

Here we use the same data as Trafalis et al. (2014). It is the dataset of meteorological parameters, circulations and observations calculated by application of the Mesocyclone Detection Algorithm (MDA) and near-storm environment (NSE) algorithm from Doppler radar velocity data. These two methods were developed at the National Severe Storms Laboratory [Kingfield et al., 2012]. The main idea of these methods is an analysis of azimuthal shear in Doppler radar velocity and rotational strength data in 3 dimensions.

In this dataset we have 10,816 observed circulations (721 of them are tornadic circulations) taken from 111 storm days. Unfortunately, these storm days are not consecutive – these circulations took place in the time period from 1995 to 1999. All observations contain 83 attributes, which describe physical characteristics like pressure, temperature, wind velocity, etc. In addition, one parameter shows the date of circulation occurrence (month). In turn, the target parameter is binary (tornado or non-tornado), which means that our problem is classification of these data into two classes.

Additionally, it is important to highlight some specific features of these data which interfere with tornado prediction. For instance, some observations account for one air circulation. The difference between them is explained by the fact that one mesocyclone was detected at different times by different radars.

Because, there is an assumption that the parameters leading to tornadogenesis are location invariant, a feature of the dataset is the absence of information about the place, where observation was made. Thus, we have neither information about any characteristics of observation location (hill existence or distance from the ocean, for instance), nor just approximate locality (for example, state).

### 3.2. Data correction

Any data mining process requires data preprocessing, such as deleting duplicate or outlying values and verification of value intervals. In addition, in some situations it can be important to check measure units.

Thus, we began examining the data in terms of outliers. For instance, attribute V23 (meso low-level convergence) for one circulation had a value equal to –32,768. In meteorological data sets, this undocumented value is normally taken as a flag for missing data. However, from the theoretical point of view the range of values for this parameter is from 0 to 70. Therefore, this observation was replaced by the mean value of the parameter.[2]

Afterwards we analyzed the data in terms of intervals of parameter values. Investigating all attributes one by one we got first hint for possible problems at V2 (meso base). The problem is that all values of the parameter lay from 0 to 229, although in theory they can reach the level of 12,000. Nevertheless, it is only a potential hint, because these data might be a real situation with a small range of the values takes place.

---

[2] At the next step we will use a better methodology to replace outliers, which is data imputation with machine learning algorithms [Richman et al., 2008].

However, studying V4 (meso strength rank) we find one more hint for incorrect data. Namely, the values should be from 0 to 25, while in the dataset the range of the values is much wider – from 14 to 12,337. In addition, for V5 (meso low-level diameter) we can find a similar situation: theoretical range (0–15,000) and the real one (0–18) differ strongly from each other. Thereby we have the situation illustrated by Table 1. All discovered features allow us to propose the following hypothesis: parameters V3, V4, V5 should be shifted.

*Table 1.* Ranges of the attribute's values

| Attribute | V2 | V3 | V4 | V5 | V6 | V7 |
|---|---|---|---|---|---|---|
| Theoretical range | 0–12,000 | 0–13,000 | 0–25 | 0–15,000 | 0–15,000 | 0–12,000 |
| Real range | 0–229 | 8–8,453 | 14–12,337 | 0–18 | 500–14,985 | 503–14,992 |

Thus, we may notice that the values of parameters V3, V4, V5, V6, V7 suit much better to the left neighbor attribute. The rest of the study is selecting the whole list of parameters, which need shift. For this purpose, we will continue to study ranges of the attribute values.

The last parameter is V25 (meso range), is suspicious , as the values of the next attribute (from 7,628 to 10,136) do not match; because the range of the mesocyclone equal to 8,000 km seems to be excessive. In turn, V2 values, which are not used now (they should be shifted to the V1, but that attribute has the right values), lie from 0 to 229. It means that these figures suit to V25.

So, the next data alteration is the following data shift: the values for V3 – V25 attributes are shifted to the previous attributes (for instance, V3 values are now V2 values). Meanwhile, V2 values go to V25.

At the next stage we noticed another specific data feature. V51 (magnitude of the low-level shear vector (surface to 3 km above ground level)) values are equal to the values of V66 (Average Relative Humidity in 0–1 km layer) for the majority of the circulations (5,171 of 5,409). Meanwhile, for the other 238 circulations V51 values are equal to the values of V67 (0–1 km shear magnitude).

*Table 2.* V51, V66-V74 attribute values

| V51 | V66 | V67 | V68 | V69 | V70 | V71 | V72 | V73 | V74 |
|---|---|---|---|---|---|---|---|---|---|
| 393 | 393 | 603 | 352 | 412 | 131 | 175 | 605 | 3446 | 153 |
| 314 | 314 | 597 | 300 | 361 | 127 | 139 | 661 | 3955 | 160 |
| 326 | 326 | 597 | 309 | 368 | 127 | 143 | 664 | 3806 | 161 |
| 320 | 320 | 599 | 305 | 367 | 125 | 142 | 670 | 3944 | 161 |
| 381 | 381 | 607 | 347 | 410 | 129 | 183 | 643 | 3496 | 163 |
| 441 | 441 | 594 | 372 | 431 | 144 | 197 | 481 | 3286 | 116 |
| 383 | 383 | 606 | 348 | 410 | 128 | 175 | 633 | 3482 | 161 |
| 331 | 331 | 597 | 313 | 371 | 128 | 145 | 661 | 3714 | 161 |
| **384** | **86** | **384** | **607** | **348** | **411** | **133** | **193** | **643** | **3497** |
| 394 | 394 | 604 | 352 | 414 | 134 | 186 | 610 | 3453 | 153 |
| 289 | 289 | 676 | 277 | 344 | 104 | 108 | 674 | 4892 | 123 |
| 325 | 325 | 597 | 309 | 368 | 127 | 143 | 665 | 3822 | 161 |
| 390 | 390 | 605 | 351 | 413 | 133 | 187 | 621 | 3467 | 156 |
| 416 | 416 | 601 | 361 | 423 | 142 | 207 | 560 | 3390 | 137 |

As an example, we present some values from the dataset (Table 2). In this example we have circulations with V51 equal to V66 and V67 (the latter is highlighted bold). It is important to point out that highlighted circulation has different values from all other data for attributes in the range V59–V76.

This data feature allows us to propose one more hypothesis; we need another shift in the dataset. However, we should decide what piece of data will be altered; either figures for 238 circulations or attribute values for 5,171 circulations. For this purpose, we study the attributes from the physical point of view.

The best parameter for this task is V66 (average relative humidity in 0–1 km layer). In the case of shifting 238 circulations' figures, this attribute's values will be up to 632, which is certainly extremely high value for this parameter. Because of it, we decided to shift the majority of circulations (V59–V76 will be shifted to the attribute with the higher number).

However, this data shift has crucial difference from the previous one. In the previous case V2 data went to V25. However, V77 values do not go to V59. The reason for this peculiarity is the fact that the range of V77 is equal to [−13,917; 20,170].

At the last stage of preliminary data analysis we study the values of the attributes in terms of the meaning of the circulation parameters. Here we are faced with some problems again. For instance, V26 (actual surface pressure) has values varying from 7,628 mb to 10,136 mb. Obviously, the decimal before the last digit is missing as taking into account the measure unit (1 *mb* = = 0.001 *b* ≈ 0.001 *atm*) that would be impossibly high pressure, up to 10 atmospheres. According to the same principle, we point out the list of parameters with "strange" range of the values. However, the decimal can be inserted and the problem will be eliminated.

### *3.3. A correlation of the parameters*

When data preprocessing has been completed, we start data analysis. However, here we face another problem. Certainly, a large amount of data can give us a lot of information, which can help to predict a tornado. However, using of too much information may make our model excessively complicated. So, for the purpose of decreasing the amount of information analyzed, we have decided to examine the data more thoroughly and to choose parameters which will play greater role in our tornado prediction model.

For this purpose, we examined the correlation between different parameters to detect attributes with high correlation. These characteristics can be discarded, because we can predict the value of one parameter using the value of the other one. For instance, attributes V61 (Average Mixing Ratio in 0–1km layer) and V62 (Average Mixing Ratio in 0–3km layer) have correlation equal to 0,984. Taking into account this fact, we decided to use only one of these parameters.

Thus, in Table 3 we can see all pairs of attributes for which the absolute value of correlation coefficient is higher than 0,85. From each pair we discarded one of the attributes[3]. For this purpose, for each pair we studied which of the parameters belongs to a larger number of highly correlated pairs and rejected it.

It is important to mention that during parameter rejection we found not only pairs, but also triplets of attributes with high absolute value of correlation coefficient. For instance, one of such triplets consists of V2 (Meso base), V3 (Meso depth) and V18 (Meso core depth) and, according to the attribute correlation, we kept only one attribute from this triplet.

---

[3] Another option is forming a new composite variable by linear combination.

*Table 3.* Pairs of attributes with absolute value of correlation over than 0,85

| Attribute 1 | Attribute 2 | Correlation |
|---|---|---|
| V61 | V62 | 0,984 |
| V46 | V56 | 0,94 |
| V3 | V18 | 0,939 |
| V2 | V3 | 0,939 |
| V35 | V36 | 0,921 |
| V67 | V69 | 0,911 |
| V34 | V57 | 0,9 |
| V2 | V18 | 0,898 |
| V69 | V75 | 0,897 |
| V51 | V68 | 0,882 |
| V11 | V17 | 0,87 |
| V4 | V19 | 0,869 |
| V40 | V43 | 0,859 |
| V31 | V63 | 0,859 |
| V37 | V41 | −0,862 |
| V40 | V41 | −0,864 |
| V50 | V54 | −0,874 |
| V34 | V58 | −0,933 |
| V57 | V58 | −0,935 |

## 3.4. An analysis of the distribution of parameters

The next step of choosing characteristics consists of choosing parameters which have different distribution of the tornadic and non-tornadic circulations. This procedure helps us to detect tornadoes in the following way. If a parameter has different distribution for circulations of different types, it has ranges, where tornadic circulations occur more rarely. As a result, we can be more confident that certain circulation is nontornadic.

For instance, on Figure 1 we can see the distribution of tornadic and non-tornadic circulations according to the parameter V3 (meso depth of the circu-lation). For this parameter, if an element has a small value in the range 0–1,000 m, we can claim that it does not seem to be a tornado. Meanwhile, if the value is in the range 10,000–12,000 m, the circulation has higher possibility to become tornado.

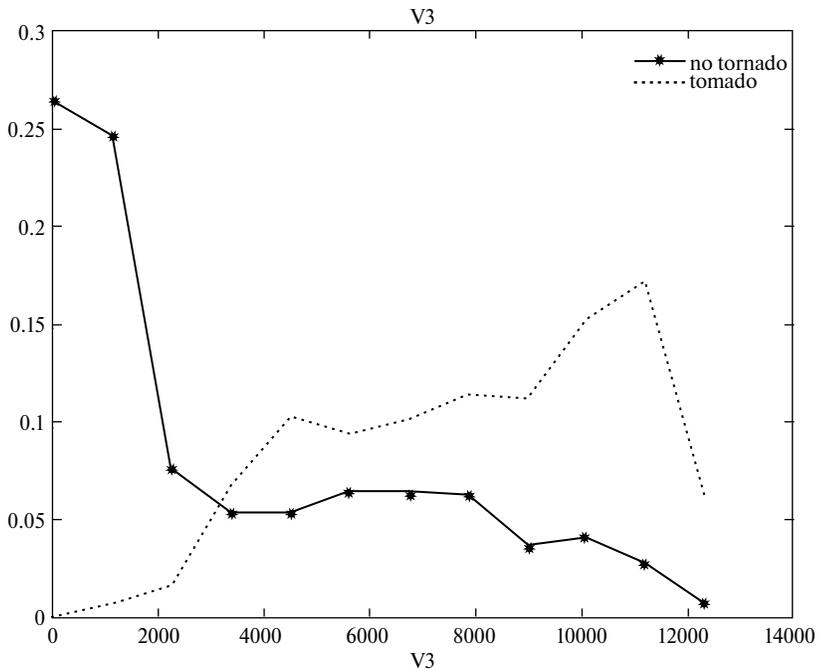

**Fig. 1.** Distribution of tornadic and non-tornadic circulation according to V3 (meso depth)

### 3.5. Additional considerations

The next step will help us to add attributes of circulation, which have not been highlighted. These parameters can be important in tornado evolution process, so we studied the procedure of tornadic circulation formation paying attention to the physical processes.

At this stage we should remember that the main goal is prediction of tornados in advance, to have enough lead time to warn people. Therefore, only those selected parameters that aid in this goal should be used in our research.

To that end, we studied the tornadogenesis process from the physical point of view. As a result, we divided the list of all parameters into two parts: *a priori* characteristics, which give a signal for tornado occurrence beforehand, and *a posteriori* ones, which only fix the prediction when the disaster has already happened. For instance, such attribute as meso low-level rotational velocity might be good *a priori* predictor of tornadogenesis. However, in these data, we did not find evidence, that this is the case and excluded it. Afterwards choosing only *a priori* parameters gives us the final list of characteristics used in finding patterns.

## 4. Framework

Consider a finite set $A$ of alternatives evaluated by $n$ parameters, i.e. the vector of values $(u_1(x), \ldots, u_n(x))$ is assigned to each alternative $x$ from $A$, i.e.,

$$x \in A \rightarrow (u_1(x), \ldots, u_n(x))$$

The problem lies in constructing a transformation $C(\cdot)$ – the rule of aggregation over $A$ – such that $C(\cdot): A \rightarrow R^1$.

Our model is based on a superposition principle, applied for different choice functions. There are different ways on how to construct a choice function $C(\cdot)$. One of the options is provided below:

$$\forall X \subseteq A\, C(X) = \{y \in X | \alpha_i u_i(y) + \alpha_j u_j(y) \geq b\},$$

where $u_i(\cdot)$ and $u_j(\cdot)$ are the values of two exact parameters $i, j \in \{1, \ldots, n\}$ of an observation, while $b$ is the threshold value that depends on the initial set $X$ and chosen parameters $i$ and $j$. In turn the values $\alpha_i$, $\alpha_j$ are automatically defined by the following linear optimization problem

$$\begin{cases} \min_{\alpha_i, \alpha_j} \sum_{l=1}^{N} e_l \\ e_l = \sqrt{(t_l - \hat{t}_l)^2} \, \forall l = 1, \ldots, N \\ \hat{t}_l = \alpha_i u_i(y_l) + \alpha_j u_j(y_l) \, \forall l = 1, \ldots, N, \end{cases}$$

where $t_l$ is the variable, which shows the ground truth of *l*-th observation (tornado or no tornado); *N* – number of observations in the dataset *X*.

Note that other forms of the choice function can be used in the model.

As a result, we got a sequential pointing out group of tornadic observations (Fig. 2).

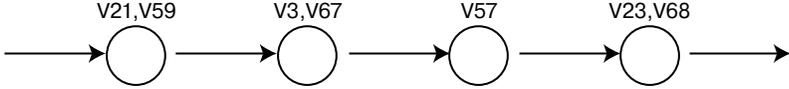

**Fig. 2.** Example of choice functions' superposition
(inscriptions above the vertices present the parameters used at this stage)

The main idea of our model is the construction of a certain number of such superposition sequences in order to distinct tornadic and non-tornadic observations. Afterwards, we unite them and the resulting prediction of our model will be the union of observation sets, obtained by different superpositions.

## 5. Application of the model

As it was mentioned before, we apply our model to the preprocessed data obtained from the University of Oklahoma. For the correct application we need to construct training and testing datasets. According to the standard rules of data mining the whole dataset should be divided into two parts in the ratio 70:30, where the larger part is the training set.

However, here we face the following problem. At the end of the study we will compare our model with the previous ones. In their study Trafalis et al. [2014] used another separation ratio – they divided the original dataset into two equal parts. As a result, if we use the standard ratio, the comparison would not be correct, because increasing the size of training dataset improves the accuracy of the model. Thus, we decided to apply the model twice. The first application will use equal separation and help us to compare the results. In turn, the second application will use standard data division ratio and show the accuracy of the model in standard conditions.

After the data sampling, it is important to choose the metrics for model comparison/evaluation, because there are numerous available "efficiency calculations." In Table 4 four of them are presented.

*Table 4.* Different efficiency calculation methods.

*tp* − true positive prediction, *fp* – false positive prediction, *tn* – true negative prediction, *fn* – false negative prediction

| Efficiency calculation method | Formula |
|---|---|
| Classification Accuracy | $CA = \dfrac{tp + tn}{tp + tn + fp + fn}$ |
| Probability of Detection | $POD = \dfrac{tp}{tp + fn}$ |
| False Alarm Ratio | $FAR = \dfrac{fp}{tp + fp}$ |
| Critical Success Index | $CSI = \dfrac{tp}{tp + fp + fn}$ |

Analyzing the efficiency calculation methods, at the first stage we reject Probability of Detection and False Alarm Ratio, because any one of these techniques does not take into account significant part of the results. For instance, the former takes into consideration only predictions for tornadic circulations. In turn, the latter does not consider the number of missed tornadoes. Two methods would need to be examined to obtain a characterization of the model.

At the second stage we should choose between Classification Accuracy and Critical Success Index. Given the peculiarity of the data, wherein about 93% of observations are non-tornadic, we reject Classification Accuracy since if we use classification accuracy all models will have almost the same results. Thus, we choose Critical Success Index.

When all preparations are over we apply the constructed tornado prediction model and evaluate its efficiency. As a result, we obtained the following figures (Table 5) for 50:50 division ratio (the tornado prediction procedure was repeated 20 times and the value in Table 5.2 is the mean value).

In turn, in case of standard 70:30 data division for training and testing sets the efficiency of our model is even higher and equal to 0,63.

Comparing all tornado prediction methods in terms of three techniques of efficiency evaluation (POD, FAR and CSI), we will get the following Roebber's (2009) performance diagram (Fig. 3). As we can see, the improved de-cision tree Pareto-dominates all other models except slight predominance of logistic regression in terms of probability of detection.

*Table 5.* Comparison of the constructed model with previous works.

| Prediction technique | POD | FAR | CSI |
|---|---|---|---|
| SVM | 0,68 | 0,22 | 0,57 |
| Logistic regression | 0,7 | 0,25 | 0,57 |
| RANF | 0,58 | 0,17 | 0,51 |
| ROTF | 0,61 | 0,21 | 0,53 |
| Superposition with mixed parameters | 0,68 | 0,16 | 0,61 |

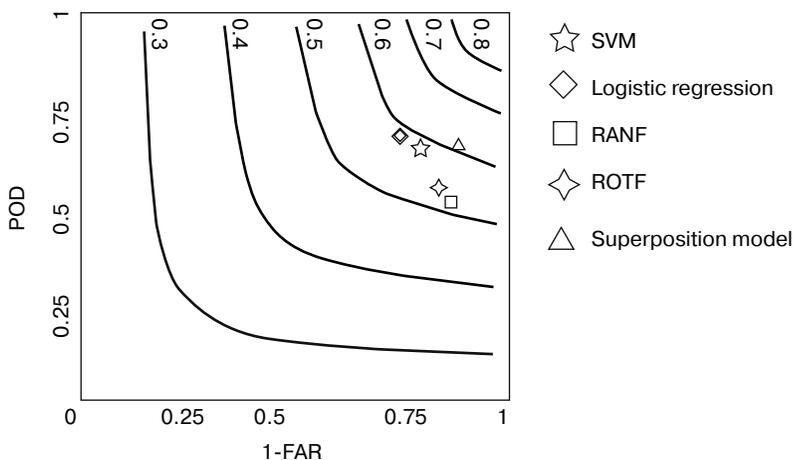

**Fig. 3.** Roebber's performance diagram for the performance results of all classifiers. The solid lines represent CSI

## 6. Conclusion

As expected, the results of improved decision tree model exceed the results of all models, which were applied before. The improvement arises as the false alarm ratio is notably smaller with this technique. There is a small trade-off of impressed misses, compared to logistic regression. According to mete-orologists' opinion, a skill improvement equal to 0,05 is significant development. As CSI is an important component of skill, constructed model is a successful attempt to improve tornado prediction technique. Therefore, the improved decision tree model provides the best opportunity to provide accurate tornado formation predictions with sufficient lead times.

# Appendix 1 – list of attributes

| Attribute number | Meaning of the attribute |
|---|---|
| V1 | Month |
| V2 | Meso base (m) [0–12000] |
| V3 | Meso depth (m) [0–13000] |
| V4 | Meso strength rank [0–25] |
| V5 | Meso low-level diameter (m) [0–15000] |
| V6 | Meso maximum diameter (m) [0–15000] |
| V7 | Meso height of maximum diameter (m) [0–12000] |
| V8 | Meso low-level rotational velocity (m/s) [0–65] |
| V9 | Meso maximum rotational velocity (m/s) [0–65] |
| V10 | Meso height of maximum rotational velocity (m) [0–12000] |
| V11 | Meso low-level shear (m/s/km) [0–175] |
| V12 | Meso maximum shear (m/s/km) [0–175] |
| V13 | Meso height of maximum shear (m) [0–12000] |
| V14 | Meso low-level gate-to-gate velocity difference (m/s) [0–130] |
| V15 | Meso maximum gate-to-gate velocity difference (m/s) [0–130] |
| V16 | Meso height of maximum gate–to–gate velocity difference (m) [0–12000] |
| V17 | Meso core base (m) [0–12000] |
| V18 | Meso core depth (m) [0–9000] |
| V19 | Meso age (min) [0–200] |
| V20 | Meso strength index (MSI) weighted by average density of integrated lyr [0–13000] |
| V21 | Meso strength index (MSIr) "rank" [0–25] |

| Attribute number | Meaning of the attribute |
|---|---|
| V22 | Meso relative depth (%) [0–100] |
| V23 | Meso low-level convergence (m/s) [0–70] |
| V24 | Meso mid-level convergence (m/s) [0–70] |
| V25 | Meso Range (km) [0–230] |
| V26 | Actual surface pressure (mb) |
| V27 | Height of the 273 K temperature surface (m agl) |
| V28 | Average wind speed over a specified depth (m/s) |
| V29 | U – component of estimated storm motion vector (north–relative; m/s) |
| V30 | V – component of estimated storm motion vector (north –relative; m/s) |
| V31 | Estimated 0–3 km storm relative helicity ($m^2/s^2$) |
| V32 | Surface relative humidity (percent) |
| V33 | Surface virtual temperature (Kelvin) |
| V34 | Downdraft CAPE (dCAPE) for a parcel 1 km AGL |
| V35 | dCAPE for a parcel 3 km AGL |
| V36 | dCAPE for the parcel at 0 Celsius |
| V37 | Convective Available Potential Energy (CAPE) (J/kg) for a parcel with Average characteristics over the lowest X mb (e.g., 100 mb) |
| V38 | Convective Available Potential Energy (CAPE) (J/kg) for Convective Inhibition (CIN) |
| V39 | Same as 37, except for Level of Free Convection (LFC) |
| V40 | Same as 37, except for Equilibrium Level (EL) |
| V41 | Same as 37, except for Lifted Index (LI) |
| V42 | Same as 37, except for Energy-Helicity Index (EHI) |
| V43 | Average Parcel Level: level (m AGL) at which the negative area above the Surface-based CAPE counterbalances the surface–based CAPE. |
| V44 | Magnitude of the storm–relative flow for the 0–2 km agl layer (kts). The estimated storm motion vector is subtracted from the average model wind vector of the layer. |
| V45 | Same as 44, except for the 4–6 km agl layer |
| V46 | Same as 44, except for the 9–11 km agl layer |

| Attribute number | Meaning of the attribute |
|---|---|
| V47 | Bulk Richardson Number (BRN) calculated according to Equations 3.4.57 and 3.4.58 of Bluestein (Vol. II). The surface CAPE is used in the calculation. Also, because of the volatility of BRN, I take log base 10 of the BRN for display purposes. |
| V48 | BRN shear (kts^2). See Eq. 3.4.58 of Bluestein Vol. II. |
| V49 | Bulk Richardson Number (BRN), except for the most unstable parcel. |
| V50 | Temperature difference (C) between 700 and 500 mb. |
| V51 | Magnitude (kts) of the low-level shear vector (surface to 3 km agl) |
| V52 | Average wind speed in the 500–300 mb layer (knots) |
| V53 | Maximum theta-e (Kelvin) in the lowest 300 mb |
| V54 | Mean lapse rate in the 850–500 mb layer (C/km) |
| V55 | Mean shear through a specified depth (hodograph length over a specified depth divided by that depth; *1000 s^–1) |
| V56 | Deep–layer shear vector magnitude (knots). The upper vector is the 9–11 km mean wind vector and the lower vector is the 0–2 km mean wind vector. |
| V57 | Average parcel LCL (m agl). See NOTE above. |
| V58 | Average RH (percent) below the surface parcel's LCL |
| V59 | Average RH (percent) below the average parcel's LCL |
| V60 | Vorticity Generation Potential (VGP) using the surface–based CAPE |
| V61 | Average Mixing Ratio in 0–1 km layer (g/kg) |
| V62 | Average Mixing Ratio in 0–3 km layer (g/kg) |
| V63 | Average Mixing Ratio in 0–6 km layer (g/kg) |
| V64 | Estimated 0–1 km storm relative helicity (m^2/s^2) |
| V65 | Estimated 0–2 km storm relative helicity (m^2/s^2) |
| V66 | Average Relative Humidity in 0–1 km layer (%) |
| V67 | 0–1 km shear magnitude (knots) |
| V68 | 0–3 km shear magnitude (knots) |
| V69 | 27% most-unstable parcel EL (corresp. to 0–3 km) shear magnitude (knots) |
| V70 | 55% most-unstable parcel EL (corresp. to 0–6 km) shear magnitude (knots) |
| V71 | 0–18% most-unstable parcel EL (corresp. to 0–2 km) storm–relative flow (knots) |

| Attribute number | Meaning of the attribute |
|---|---|
| V72 | 36–55% most-unstable parcel EL (corresp. to 4–6 km) storm–relative flow (knots) |
| V73 | 82–100% most-unstable parcel EL (corresp. to 9–11 km) storm–relative flow (knots) |
| V74 | Level of Maximum Buoyancy from most-unstable parcel (m AGL) |
| V75 | Maximum Buoyancy (from most-unstable parcel) (m2/s2) |
| V76 | Level of Maximum Buoyancy from most-unstable parcel (corresp. to 0–6 km) shear magnitude (knots) |
| V77 | 20% below Level of Maximum Buoyancy from most-unstable parcel (corresp. to 4–6 km) storm–relative flow (knots) |
| V78 | Low–Level (Most–unstable–parcel LFC to Most-unstable-parcel LFC+1km) Lapse Rate (C/km) |
| V79 | Normalized most–unstable parcel CAPE [divide by z (Most-unstable parcel EL) – z (Most-unstable parcel LFC)] |
| V80 | Most-unstable parcel CAPE Most–unstable parcel CAPE from Most-unstable parcel LFC to Most-unstable parcel LFC+3km |
| V81 | Most–unstable parcel CAPE from Most-unstable parcel LFC to Most-unstable parcel LFC+3km divided by Most-unstable parcel CAPE (% age of total) |
| V82 | Most-unstable parcel CAPE from surface to 3 km AGL |
| V83 | Most-unstable parcel CAPE from surface to 3 km AGL divided by Most-unstable parcel CAPE (% age of total) |

Изучены существующие методы предсказания торнадо, а также основные механизмы возникновения этих природных катастроф. Построена модель, основанная на методе суперпозиции. Для проверки эффективности построенная модель была протестирована на реальных данных, полученных из Университета Оклахомы (США).

Предложенная модель предсказания торнадо значительно превосходит предыдущие методы с точки зрения эффективности.

*Алескеров Фуад* – Национальный исследовательский университет «Высшая школа экономики», Институт проблем управления Российской академии наук им. В.А. Трапезникова, Москва, Российская Федерация.

*Байбородов Никита* – Московский физико-технический институт, Москва, Российская Федерация.

*Демин Сергей* – Национальный исследовательский университет «Высшая школа экономики», Москва, Российская Федерация.

*Ричман Майкл* – Университет Оклахомы, США.

*Трафалис Теодор* – Университет Оклахомы, США.

*Швыдун Сергей* – Национальный исследовательский университет «Высшая школа экономики», Институт проблем управления Российской академии наук им. В.А. Трапезникова, Москва, Российская Федерация.

*Якуба Вячеслав* – Национальный исследовательский университет «Высшая школа экономики», Институт проблем управления Российской академии наук им. В.А. Трапезникова, Москва, Российская Федерация.

Алескеров Фуад, Байбородов Никита, Демин Сергей, Ричман Майкл, Трафалис Теодор, Швыдун Сергей, Якуба Вячеслав

# Построение эффективной модели машинного обучения для предсказания торнадо

(*на английском языке*)